\documentclass[aps,preprint,floats,epsf,epsfig,nofootinbib,letter]{revtex4}

\usepackage{color}
\usepackage{graphicx}
\usepackage{dcolumn}
\usepackage{bm}

\def\be{\begin{eqnarray}}
\def\en{\end{eqnarray}}
\def\non{\nonumber}

\def\la{\langle}
\def\ra{\rangle}

\begin{document}

\begin{flushright}
\end{flushright}

\title{Revisiting Scalar and Pseudoscalar Couplings with Nucleons}

\author{ Hai-Yang Cheng}
\affiliation{Institute of Physics, Academia Sinica, Taipei, Taiwan 11529, ROC}

\author{ Cheng-Wei Chiang}
\affiliation{Department of Physics and Center for Mathematics and Theoretical Physics,
National Central University, Chungli, Taiwan 32001, ROC}
\affiliation{Institute of Physics, Academia Sinica, Taipei, Taiwan 11529, ROC}
\affiliation{Physics Division, National Center for Theoretical Sciences, Hsinchu, Taiwan 30013, ROC}

\bigskip
\begin{abstract}
\bigskip
Certain dark matter interactions with nuclei are mediated possibly by a scalar or pseudoscalar Higgs boson.  The estimation of the corresponding cross sections requires a correct evaluation of the couplings between the scalar or pseudoscalar Higgs boson and the nucleons.  Progress has been made in two aspects relevant to this study in the past few years.  First, recent lattice calculations show that the strange-quark sigma term $\sigma_s$ and the strange-quark content in the nucleon are much smaller than what are expected previously.  Second, lattice and model analyses imply sizable SU(3) breaking effects in the determination on the axial-vector coupling constant $g_A^8$ that in turn affect the extraction of the isosinglet coupling $g_A^0$ and the strange quark spin component $\Delta s$ from polarized deep inelastic scattering experiments.
Based on these new developments, we re-evaluate the relevant nucleon matrix elements and compute the scalar and pseudoscalar couplings of the proton and neutron.  We also find that the strange quark contribution in both types of couplings is smaller than previously thought.

\end{abstract}



\maketitle
\small
%
\section{Introduction \label{sec:intro}}
%

Knowledge of scalar and pseudoscalar Higgs boson interactions with the nucleons at low energies is an important ingredient for computing certain dark matter--nuclei interaction cross sections.  For example, the scalar interactions between fermionic dark matter and nucleons mediated by the scalar Higgs boson can be studied by measuring spin-independent (SI) cross sections, while interactions of the dark matter with the nucleons mediated by light pseudoscalars can be probed by studying the momentum-dependent, spin-dependent (SD) cross sections. \footnote{Since the cross sections for the dark matter interacting with the nucleons induced by pseudoscalars vanish in the non-relativistic limit, it means that the signature of dark matter cannot be seen by virtue of SD direct detection.}
At the quark level, the interactions of a neutral scalar $\phi$ and a pseudoscalar $\sigma$ with quarks have the general expressions:
\be
{\cal L}_{\phi qq} &=& (\sqrt{2}G_F)^{1/2}\sum_q \zeta_q m_q \bar qq\phi, \non \\
{\cal L}_{\sigma qq} &=& (\sqrt{2}G_F)^{1/2}\sum_q \xi_q m_q \bar qi\gamma_5q\phi,
\en
where $\zeta_q$ and $\xi_q$ are the couplings of the scalar $\phi$ and the pseudoscalar $\sigma$ with quarks, respectively. While in the standard model $\zeta_q=1$ for all $q$'s, they may have values different from unity beyond the standard model.  To evaluate the effective scalar and pseudoscalar couplings with the nucleons given by
\be
g_{\phi\! N\!N} &=& (\sqrt{2}G_F)^{1/2}\sum _q\la N|\zeta_q m_q\bar qq|N\ra, \mbox{ and}\non \\
g_{\sigma\! N\!N} &=& (\sqrt{2}G_F)^{1/2}\sum _q\la N|\xi_q m_q\bar qi\gamma_5q|N\ra,
\en
it amounts to computing the nucleon matrix elements of quark scalar and pseudoscalar densities, namely,
$\la N|m_q\bar qq|N\ra$ and $\la N|m_q\bar qi\gamma_5 q|N\ra$.

The coupling $g_{\phi\! N\!N}$ was first studied by Shifman, Vainstein and Zakharov (SVZ) \cite{Shifman} who assumed a negligible strange quark contribution to the nucleon mass.  Consequently, their scalar-nucleon coupling was dominated by heavy quarks. Based on the pion-nucleon sigma term available in late 80's which implied a sizable strange quark content in the nucleon, it was shown in \cite{TPCheng88,Cheng1989} that the effective coupling $g_{\phi\! N\!N}$ was dominated by the $s$ quark rather than by heavy quarks.  As a result, the scalar-nucleon coupling was enhanced by a factor of about 2.5 \cite{Cheng1989}.

Although the coupling of the axion, which is an example of the pseudoscalar boson, with the nucleon had been studied before (see \cite{Cheng1988} for a review), a thorough discussion of generic pseudoscalar-nucleon couplings was only given later in \cite{Cheng1989} based on the relation
\be \label{eq:sum}
\la N|\bar ui\gamma_5u+\bar di\gamma_5d+\bar si\gamma_5s|N\ra=0\,,
\en
derived from the large-$N_c$ and chiral limits.  Contrary to the case of scalar-nucleon couplings, $g_{\sigma\! N\!N}$ does receive significant contributions from light quarks even in the chiral limit.  It is conventional to express $g_{\sigma\! N\!N}$ in terms of the axial-vector couplings $g_A^a$ $(a=0,3,8)$ or the quark spin components $\Delta q$ with $q=u,d,s$.

Recently, there is some progress in topics relevant to the interactions between the scalar or pseudoscalar Higgs boson and the nucleons.  First, there have been intensive lattice calculations of the pion-nucleon sigma term $\sigma_{\pi\!N}$, the quark sigma term $\sigma_q\equiv m_q\la p|\bar qq|p\ra$, and the strange quark content in the nucleons characterized by the parameter $y$.  Especially, the lattice calculation of $\sigma_s$ is not available until recent years.  The lattice results indicate that the strange quark fraction in the nucleon and $\sigma_s$ are getting smaller than what we had one or two decades ago.  For example, the new lattice average of $\sigma_s=(43\pm8)$ MeV \cite{Lin} is much smaller than the value of $\sim 390$ MeV in the late 80's and $\sim 130$ MeV in the early 90's.
Second, recent lattice and model calculations \cite{QCDSF:Deltas,Bass} hint at a size of $g_A^8$ about 20\% smaller than the canonical value $0.585$ determined from the hyperon $\beta$-decays supplemented by flavor SU(3) symmetry.  Since $g_A^0$ is extracted from the measurement of the first moment of the proton polarized structure function $g_1^p$, a decrease in $g_A^8$ will lead to an increase in $g_A^0$.  This in turn implies a $\Delta s$ reduced in magnitude by a factor of 2 to 3.

Motivated by the above-mentioned developments, in this work we would like to revisit the couplings of scalar and pseudoscalar bosons with the nucleons to incorporate the recent progress.  Moreover, we wish to investigate if it is possible to evaluate the matrix elements $\la N|G\tilde G|N\ra$ and $\la N|\bar qi\gamma_5 q|N\ra$ without invoking the large-$N_c$ chiral relation Eq.~(\ref{eq:sum}) as this relation is presumably subject to $1/N_c$ and chiral corrections.

This paper is organized as follows.  Section~\ref{sec:scalarHiggs} deals with the couplings between the nucleons and the scalar Higgs boson, whereas Section~\ref{sec:pseudoHiggs} analyzes the case with a pseudoscalar Higgs boson.  Our findings are summarized in Section~                                                                \ref{sec:summary}.  Appendix~\ref{sec:comparison} provides a comparison of our results against two sets of parameters commonly used in dark matter physics analyses.

%
\section{Scalar Higgs couplings to the nucleons \label{sec:scalarHiggs}}
%

The baryon mass can be expressed in terms of the matrix element of the trace of the energy-momentum tensor
\be
\Theta^\mu_{~\mu}=m_u\bar uu+m_d \bar dd+m_s\bar ss+\sum_h m_h\bar q_hq_h+{\beta\over 4\alpha_s}GG \ ,
\en
with
\be
\beta=-{b\alpha_s^2\over 2\pi}, \qquad b=11-{2\over 3}n_\ell-{2\over 3}n_h\ ,
\en
where $n_{\ell,h}$ denote the numbers of light and heavy quarks, respectively, and the trace anomaly is governed by the $GG \equiv G_{\mu\nu}^aG^{a\mu\nu}$ term.  As pointed out by Shifman, Vainstein and Zakharov~\cite{Shifman}, heavy quarks contribute to the baryon mass via the triangle diagram with external gluons.  Under the heavy quark expansion~\cite{Shifman}
\be \label{eq:HQE}
m_h\bar q_hq_h\to -{2\over 3}\,{\alpha_s\over 8\pi}GG+{\cal O}\left({\mu^2\over m_h^2}\right) \ ,
\en
where $\mu$ is a typical hadron mass scale. Thus we have
\be \label{eq:trace}
\Theta^\mu_{~\mu}=m_u\bar uu+m_d \bar dd+m_s\bar ss -{9\alpha_s\over 8\pi}GG.
\en

Under the light quark mass expansion, the baryon mass has the general expression (for a review, see \cite{Scherer})
\be \label{eq:expansion}
m_{\cal B}=m_0+\sum_qb_qm_q+\sum_qc_qm_q^{3/2}+\sum_q d_qm_q^2\,{\rm ln}{m_q\over m_0}+\sum_qe_qm_q^2+{\cal O}(m_q^3) ~,
\en
where the coefficients $b_q$, $c_q$, $d_q$ and $e_q$ are baryon state-dependent. The $m_q^{3/2}$ terms arise from chiral loop contributions which are ultraviolet finite and non-analytic in the quark masses as $m_P^3\propto m_q^{3/2}$ with $m_P$ being the pseudoscalar meson mass in the loop \cite{Gasser1981,Gasser1991,Jenkins,Bernard:1993}. Terms of order $m_q^2$ have been calculated in~\cite{Borasoy,Borasoy:1996}, while terms up to $m_q^3$ can be found in~\cite{Schindler}. Defining the scale-independent quark sigma term of the proton
$\sigma_q\equiv m_q\la p|\bar qq|p\ra$,
it can be determined from Eq.~(\ref{eq:expansion}) through the relation
\be
\sigma_q=m_q{\partial m_p\over \partial m_q}\ .
\en
This relation can be derived by the aid of the Feynman-Hellmann theorem~\cite{Gasser:1987}.

Using the proton matrix elements defined by
\be
A\equiv -\left< p \left|{9\alpha_s\over 8\pi}GG \right|p \right>, \quad B_u\equiv \la p|\bar uu|p\ra, \quad B_d\equiv \la p|\bar dd|p\ra, \quad B_s\equiv \la p|\bar ss|p\ra
\en
and applying Eq.~(\ref{eq:trace}) and the flavor SU(3) symmetry,
the masses of the octet baryons can be expressed as \cite{Gasser1982}
\be \label{eq:Baryonmass}
m_p &=&A+B_u m_u+B_dm_d+B_sm_s, \non \\
m_n &=& A+B_dm_u +B_um_d+B_sm_s, \non \\
m_{\Sigma^+} &=& A+B_um_u +B_sm_d+B_dm_s, \non \\
m_{\Sigma^0} &=& A+(B_u+B_s)\hat m+B_dm_s,  \\
m_{\Sigma^-} &=& A+B_sm_u +B_um_d+B_dm_s, \non \\
m_{\Xi^0} &=& A+B_dm_u +B_sm_d+B_um_s, \non \\
m_{\Xi^-} &=& A+B_sm_u +B_dm_d+B_um_s, \non \\
m_\Lambda &=& A+{1\over 3}(B_u+4B_d+B_s)\hat m+{1\over 3}(2B_u-B_d+2B_s)m_s, \non
\en
with $\hat m\equiv (m_u+m_d)/2$. We thus see that the quark mass expansion of the baryon masses to the leading order yields $A=m_0$ and $B_q=b_q$.
From the mass formulae Eq.~(\ref{eq:Baryonmass}) one can derive the Gell-Mann Okubo mass relation in the baryon sector
\be
m_{\Sigma}-m_N={1\over 2}(m_\Xi-m_N)+{3\over 4}(m_\Sigma-m_\Lambda) \ ,
\en
the Coleman-Glashow mass formula
\be
m_p-m_n+m_{\Sigma^-}-m_{\Sigma^+}+m_{\Xi^0}-m_{\Xi^-}=0 \ ,
\en
and the constraint such as
\be
&& {m_{\Sigma^+}-m_{\Sigma^-}\over m_{\Xi^0}-m_{\Xi^-}}={m_\Xi-m_N\over m_\Sigma-m_N}
\en
with $m_N\equiv(m_p+m_n)/2$, $m_\Sigma\equiv(m_{\Sigma^+}+m_{\Sigma^-})/2$ and $m_\Xi\equiv(m_{\Xi^0}+m_{\Xi^-})/2$.

Using the current quark masses \cite{PDG}
\be
m_u=2.5^{+0.6}_{-0.8}\,{\rm MeV} \ , \qquad m_d=5.0^{+0.7}_{-0.9}\,{\rm MeV} \ , \qquad
m_s=100^{+30}_{-20}\,{\rm MeV} ~,
\en
normalized at the scale $\mu=2$ GeV, we obtain
\footnote{From the mass formulae given in Eq.~(\ref{eq:Baryonmass}), we also have the relations
$B_u-B_s=(m_{\Sigma^+}-m_{\Sigma^-})/(m_u-m_d)=3.23$ and
$B_d-B_s=(m_{\Xi^0}-m_{\Xi^-})/(m_u-m_d)=2.74$. It is clear that the value of $B_u-B_s$ obtained in this manner differs sizably from that in Eq.~(\ref{eq:Bq}). Indeed, in this case one needs to take into account electromagnetic corrections to the isospin splitting.}
\be \label{eq:Bq}
B_u-B_s &=&{2(m_\Xi-m_N)\over 2m_s-m_u-m_d}=3.94 \ , \non \\
B_d-B_s &=& {2(m_\Sigma-m_N)\over 2m_s-m_u-m_d}=2.64
\en
at $\mu=2$ GeV. To determine the parameters $B_u$, $B_d$ and $B_s$ we need additional information. To proceed, we define two quantities:
\be \label{eq:y}
y\equiv {2 B_s\over B_u+B_d} ~,
\en
which characterizes the strange quark content in the proton and
\be
\sigma_0\equiv \hat m\la N|\bar uu+\bar dd-2\bar ss|N\ra ~,
\en
which is related to the pion-nucleon sigma term
\footnote{The pion-nucleon sigma term is the scalar form factor $\sigma(t)$ at $t=0$ defined by $\la N(p')|\hat m (\bar uu+\bar dd)|N(p)\ra=\bar u(p')\sigma(t)u(p)$.}
\be
\sigma_{\pi\! N}\equiv \hat m\la N|\bar uu+\bar dd|N\ra
\en
by
\be \label{eq:y&sigma}
\sigma_{\pi\! N}={\sigma_0\over 1-y} \ .
\en
Hence, $\sigma_0$ is the pion-nucleon sigma term in the absence of the strange quark content in the nucleon.
It follows from Eq.~(\ref{eq:Baryonmass}) that at the leading order
\footnote{The sigma term $\sigma_0$ can also be expressed as
\be
\sigma_0={m_d+m_u\over m_d-m_u}\,[m_{\Sigma^-}-m_{\Sigma^+}-{1\over 2}(m_n-m_p)] \non \ .
\en
However, it is necessary to take into account the electromagnetic corrections to the isospin splittings \cite{Gasser1982}. }
\be \label{eq:sigma0}
\sigma_0={3\hat m\over m_s-\hat m}\,(m_{\Xi}-m_\Lambda), \quad {\rm or}\quad
\sigma_0={\hat m\over m_s-\hat m}\,(m_{\Xi}+m_\Sigma-2m_N),
\en
Numerically, $\sigma_0=23.7$ MeV and 24.7 MeV, respectively. The experimental and/or theoretical information of $\sigma_{\pi\!N}$ will enable us to determine the strange quark fraction $y$ which puts an additional constraint on the parameters $B_u$, $B_d$ and $B_s$.

In the early study based on the sigma term $\sigma_{\pi\!N}=55\sim 60$ MeV, one has a large strange quark content $y\approx 0.55$ and
\be
\sigma_u\approx 20~{\rm MeV}, \qquad \sigma_d\approx 33~{\rm MeV}, \qquad \sigma_s\approx 394~{\rm MeV},
\en
The strange quark content $y$ and the sigma term $\sigma_s$ are both unexpectedly large.

In recent years, there exist many lattice results on the strange-quark sigma term $\sigma_s$ either based on a direct calculation \cite{Takeda:a,Takeda:b,Bali:ms,Engelhardt,Dinter} or using the Feynman-Hellman theorem through a study of how the nucleon mass varies with $m_s$ \cite{Ohki,Toussaint,Young,Durr,Horsley}.  A global fit to the current dynamical lattice data yields $\sigma_s=43\pm8$ MeV \cite{Lin}.  Meanwhile, the strange quark content also becomes smaller, $y=0.05\sim 0.20$.  The possibility of reducing the naive estimate of $\sigma_s$ by one order of magnitude has been realized by several authors \cite{Gasser1981,Gasser1991,Jenkins,Bernard:1993,Borasoy,Borasoy:1996} by considering the corrections to the baryon masses from higher order terms in the light quark mass expansion.


Based on heavy baryon chiral perturbation theory to order ${\cal O}(m_q^2)$, Borasoy and Mei\ss ner \cite{Borasoy,Borasoy:1996} obtained
\be
\sigma_{\pi \!N}=45\,{\rm MeV}, \qquad \sigma_0=36\pm7\,{\rm MeV}, \qquad y=0.21\pm0.20 \ .
\en
Notice that they did not predict $\sigma_{\pi \!N}=45\,{\rm MeV}$ rather took it as an input. Similar results with
\be
\sigma_{\pi \!N}=45\pm7\,{\rm MeV}, \qquad \sigma_0=35\pm5\,{\rm MeV}, \qquad y=0.22\pm0.10
\en
were also obtained by the ${\cal O}(m_q^{3/2})$ calculation of Gasser, Leutwyler and Sainio~\cite{Gasser1991}.
In short, the analyses by Gasser, Leutwyler and Sainio and by Borasoy and Mei\ss ner lead to $\sigma_{\pi\!N}\approx 45$ MeV, $y\approx 0.2$ and $\sigma_s\approx 130$ MeV. As mentioned
before, the recent intensive lattice activities in the calculations of $\sigma_{\pi\!N}$, $y$
and in particular $\sigma_s$ have put previous analyses on test. It is obvious that the predicted $\sigma_s$  is still too large compared to the recent lattice average $43\pm8$ MeV.
If $\sigma_s$ is fixed to be 50 MeV, then $B_s=0.5$ at $\mu=2$ GeV. From Eqs.~(\ref{eq:y}) and (\ref{eq:y&sigma}), we have
\be
\sigma_{\pi\!N}=\sigma_0+2B_s\hat m ~.
\en
Hence, a smaller $\sigma_s$ means that the difference between $\sigma_{\pi \!N}$ and $\sigma_0$ should be smaller. With $\sigma_0=36$ MeV, the parameters $\sigma_{\pi\!N}$ and $y$ are fixed to be $\sigma_{\pi\!N}=39.75$ MeV and $y=0.094$\,.  Indeed, most recent lattice calculations in Refs.~\cite{Bali:ms} and \cite{Durr} yield $\sigma_{\pi\!N}=38\pm12$ MeV and $39\pm4^{+18}_{-~7}$ MeV, respectively.  To conclude, $\sigma_s$ becomes smaller because of two reasons: a significantly larger $\sigma_0$ due to higher-order quark mass corrections to the baryon masses and a smaller $\sigma_{\pi \!N}$ and its difference with $\sigma_0$.

To determine the parameters $B_u$ and $B_d$, we shall assume that the ratio of $B_u-B_s$ and $B_d-B_s$ in Eq.~(\ref{eq:Bq}) is insensitive to the higher-order quark mass corrections even for $B_s=0.5$
\be
{B_u-B_s\over B_d-B_s}={m_\Xi-m_N\over m_\Sigma-m_N}=1.49 \ .
\en
This together with the sigma term $\sigma_{\pi\!N}=39.75$ MeV yields
$\sigma_u=15.6$ MeV and $\sigma_d=21.8$ MeV.  It should be stressed that these parameters are all correlated. For example, one cannot have a small $\sigma_s$ of order 50 MeV and a $\sigma_{\pi\!N}$ larger than 40 MeV.
Various parameters are summarized in Table~\ref{tab:sigma} with $\sigma_s$ varies from 40 MeV to 60 MeV.

The pion-nucleon sigma term also can be determined experimentally by
relating it to the measured on-shell pion-nucleon scattering amplitude $\Sigma_{\pi\!N}$ at the Cheng-Dashen point through the relation
\be \label{eq:Sigma}
\Sigma_{\pi\!N}=\sigma_{\pi\!N}(2m_\pi^2)+\Delta_R=
\sigma_{\pi\!N}(0)+\Delta_\sigma+\Delta_R ~.
\en
It was found that $\Delta_\sigma=15$ MeV \cite{Gasser1991} (see also \cite{Becher}) and $\Delta_R\approx 2$ MeV \cite{Borasoy:1996}. Since the Cheng-Dashen point $s=u=m_N^2$ and $t=2m_\pi^2$ is outside of the physical region, it is customary to rely on the dispersion relation to extract $\Sigma_{\pi\! N}$ from the pion-nucleon scattering data. The Karlsruhe result of $\Sigma_{\pi\! N}=64\pm 8$ MeV by Koch \cite{Koch} has been considered to be a benchmark in nineties as it leads to $\sigma_{\pi\!N}\sim 45$ MeV. In recent years there have existed several new extractions of $\Sigma_{\pi\!N}$: $73\pm9$ MeV by Olsson \cite{Olsson}, $79\pm7$ MeV by Pavan et al. \cite{Pavan}, $81\pm 6$ MeV by Hite et al. \cite{Hite}. After applying Eq. (\ref{eq:Sigma}) to convert $\Sigma_{\pi\!N}$ into $\sigma_{\pi\!N}$, the resultant pion-nucleon term lies in the range of 60-65 MeV. Another new analysis of the $\pi N$ scattering amplitude  without extrapolating to the unphysical region using a method based on Lorentz covariant baryon chiral perturbation theory yields $\sigma_{\pi\!N}=59\pm 7$ MeV \cite{Alarcon}. Therefore, the pion-nucleon term inferred from the $\pi N$ scattering data tends to be far above the benchmark of 45 MeV, whereas the lattice calculations push $\sigma_{\pi\!N}$ to the opposite end. This conflicting puzzle between lattice and experimental results for the pion-nucleon term remains an enigma, though some possible resolution has been proposed \cite{Camalich}.

\begin{table}[t]
\caption{Summary of parameters for three choices of $\sigma_s$.
 All the terms are in units of MeV except the dimensionless $y$ and $g_{\phi\! N\!N}$.  The
sigma term $\sigma_0$ is fixed to 36 MeV. The extraction method is described in the main text. For the neutron, $\sigma_u^{(n)}=11$ MeV and $\sigma_d^{(n)}=31$ MeV for $\sigma_s=50$ MeV. To get the numerical value of $g_{\phi\! N\!N}$ we have assumed three heavy quarks and $\zeta_q=1$.
}
\label{tab:sigma}
\begin{center}
\begin{tabular}{c c c c c c c c}
\hline\hline
$\sigma_s$~~ & ~~$\sigma_u$~~ & ~~$\sigma_d$~~ & ~~$\sigma_{\pi\!N}$~~ & ~~$\la p|{\alpha_s\over 4\pi}GG|p\ra$~~ & ~~$y$~~ & $g_{\phi\! N\!N}$ \\ \hline
60~~ & ~~15.9~~ & ~~22.3~~  & ~~40.5~~ & ~~$-187$~~ & ~~0.11~~ & $1.2 \times 10^{-3}$ \\
50~~ & ~~15.6~~ & ~~21.8~~  & ~~39.8~~ & ~~$-189$~~ & ~~0.09~~ & $1.1 \times 10^{-3}$ \\
40~~ & ~~15.4~~ & ~~21.3~~  & ~~39.0~~ & ~~$-191$~~ & ~~0.08~~ & $1.1 \times 10^{-3}$ \\
\hline\hline
\end{tabular}
\end{center}
\end{table}

We are now ready to compute the effective scalar-proton coupling strength
\be \label{eq:phipp}
g_{\phi pp} &=& (\sqrt{2}G_F)^{1/2}\left(\sum_\ell\zeta_\ell \, \sigma_\ell-{\alpha_s\over 12\pi}\la p|GG|p\ra\sum_h\zeta_h\right) \non \\
&=& (\sqrt{2}G_F)^{1/2}\left(\zeta_u\sigma_u+\zeta_d\sigma_d+\zeta_s\sigma_s+{2\over 27}(m_p-\sigma_u-\sigma_d-\sigma_s)\sum_h\zeta_h\right),
\en
where use of Eqs. (\ref{eq:HQE}) and (\ref{eq:trace}) has been made.
If light quark contributions are neglected, one will have
\be
g_{\phi\!N\!N} \approx (\sqrt{2}G_F)^{1/2}{2\over 27}\,m_N\sum_h\zeta_h \ ,
\en
which is the original SVZ result \cite{Shifman}.
Assuming three heavy quarks and $\zeta_q=1$, we then have $g_{\phi pp}= 1.1\times 10^{-3}$ from Eq. (\ref{eq:phipp}) which is smaller than the previous estimate $2.2\times 10^{-3}$ \cite{Cheng1989} by a factor of 2. Note that since the difference between $g_{\phi nn}$ and $g_{\phi pp}$ is negligible, they can be denoted collectively as $g_{\phi\! N\!N}$.  We also note in passing that the value of $g_{\phi\! N\!N}$ depends on the number of heavy quarks.  The increment is $2.6 \times 10^{-4}$ for each additional heavy quark.

%
\section{Pseudoscalar Higgs couplings to the nucleons \label{sec:pseudoHiggs}}
%

The analog of the heavy quark expansion (\ref{eq:HQE}) in the pseudoscalar sector is \cite{Shifman}
\be
m_h\bar q_hi\gamma_5q_h\to -{\alpha_s\over 8\pi}G\tilde G+{\cal O}\left({\mu^2\over m_h^2}\right) \ ,
\en
with $G\tilde G\equiv {1\over 2}\epsilon_{\mu\nu\alpha\beta}G^{a\mu\nu}G^{a\alpha\beta}$. The
pseudoscalar coupling with the nucleon then becomes
\be
g_{\sigma\! N\!N}=(\sqrt{2}G_F)^{1/2}\left(\xi_u m_uE_u+\xi_d m_d E_d+\xi_s m_s E_s-\sum_h\xi_h\left< N \left| {\alpha_s\over 8\pi}G\tilde G \right| N \right> \right) ~,
\en
where $E_q\equiv \la N|\bar qi\gamma_5q|N\ra$. The quantities $E_q$ are related to the nucleon matrix elements of axial-vector currents.
The divergences of the nucleon matrix elements of the axial-vector currents
\be \label{eq:gAs}
A_\mu^3 &=& \bar u\gamma_\mu\gamma_5u-\bar d\gamma_\mu\gamma_5d ~, \non \\
A_\mu^8 &=& \bar u\gamma_\mu\gamma_5u+\bar d\gamma_\mu\gamma_5d-2\bar s\gamma_\mu\gamma_5s ~,  \\
A_\mu^0 &=& \bar u\gamma_\mu\gamma_5u+\bar d\gamma_\mu\gamma_5d+\bar s\gamma_\mu\gamma_5s ~, \non
\en
read (we have dropped $\bar \psi i\gamma_5\psi$ for convenience)
\be \label{eq:gAEi}
g_A^3m_N &=& m_uE_u-m_dE_d ~, \non \\
g_A^8m_N &=& m_uE_u+m_dE_d-2m_sE_s ~, \\
g_A^0m_N &=& m_uE_u+m_dE_d+m_sE_s
+ 3\left< N \left|\frac{\alpha_s}{8\pi}G\tilde G \right|N\right> ~, \non
\en
and $g_A^a$ are the axial-vector form factors of $\la N|A_\mu^a|N\ra$ at $q^2=0$. Owing to the gluonic anomaly, it is clear from Eq. (\ref{eq:gAEi}) that $E_u$, $E_d$ and $E_s$ cannot be determined individually without additional information on the matrix element $\la N|G\tilde G|N\ra$. In the OPE approach, the proton matrix element of the axial-vector current is related to the quark spin component: \footnote{In the late 80's and early 90's there had been a hot debate whether or not there was an anomalous gluon contribution to $\la p|\bar q\gamma_\mu\gamma_5 q|p\ra$ and to the first moment of the polarized structure function $g_1^p$ through axial anomaly.
It was realized later that it depended on the factorization scheme in defining the quark spin density $\Delta q(x)$ (for a review, see \cite{Cheng1996}). Eq. (\ref{eq:deltaq}) is valid in the gauge-invariant factorization scheme such as the $\overline{\rm MS}$ scheme.}
\be \label{eq:deltaq}
\la p|\bar q\gamma_\mu\gamma_5 q|p\ra s^\mu=\Delta q ~,
\en
where $s^\mu$ is the proton spin 4-vector and $\Delta q$ represents the net helicity of the quark flavor $q$ along the direction of the proton spin in the infinite momentum frame
\be
\Delta q=\int^1_0 \Delta q(x) dx\equiv \int^1_0\left[q^\uparrow(x)+\bar q^\uparrow(x)-q^\downarrow(x)-\bar q^\downarrow (x)\right]dx \ .
\en
The axial coupling constants can be expressed as
\be \label{eq:gA Deltaq}
g_A^3(Q^2) &=& \Delta u(Q^2)-\Delta d(Q^2) ~, \non \\
g_A^8(Q^2) &=& \Delta u(Q^2)+\Delta d(Q^2)-2\Delta s(Q^2) ~,  \\
g_A^0(Q^2) &=& \Delta u(Q^2)+\Delta d(Q^2)+\Delta s(Q^2) ~, \non
\en
with $Q^2=-q^2$. While the non-singlet couplings $g_A^3$ and $g_A^8$ are readily determined from low-energy neutron and hyperon beta decays, the singlet coupling $g_A^0$ can be inferred from experiments on polarized deep inelastic lepton-nucleon scattering.  The measured first moment of the polarized proton structure function $g_1^p(x)$ is related to $g_A^a$ via
\be \label{eq:g1p}
\Gamma_1^p(Q^2)\equiv \int^1_0 g_1^p(x,Q^2)dx=C_{\rm NS}(Q^2)\left({1\over 12}g_A^3+{1\over 36}g_A^8\right)+{1\over 9} C_{\rm S}(Q^2)g_A^0(Q^2) ~,
\en
where $C_{\rm NS}$ and $C_{\rm S}$ are the perturbative QCD corrections to non-singlet and singlet parts of $\Gamma_1^p$, respectively. They have been calculated to order $\alpha_s^3$, and the explicit expressions can be found in \cite{Larin}.

Historically, there had been several attempts for solving $E_q$'s from Eq. (\ref{eq:gAEi}):
\begin{enumerate}
 \item
Inspired by the early EMC result of a small $g_A^0$ \cite{EMC}, it was natural to consider the large-$N_c$ chiral limit ({\it i.e.}, the double limits $N_c\to\infty$ and $m_q\to 0$) \cite{Brodsky,Wakamatsu} where $g_A^0=0$ and \footnote{The relation (\ref{eq:sum1}) had also been obtained in \cite{Riazuddin} based on a different approach. For the SU(2) version of Eq. (\ref{eq:sum1}), namely, $\la p|\bar ui\gamma_5u+\bar di\gamma_5d|p\ra=0$, see \cite{Anselm}.}
\be \label{eq:sum1}
 E_u+E_d+E_s=0 \ .
\en
However, later deep inelastic scattering experiments showed that $g_A^0$ was of order 0.30 or even bigger. This means that the relations $g_A\approx 0$ and Eq.~(\ref{eq:sum1}) are subject to $1/N_c$ and chiral corrections. A study of chiral and SU(3) breaking corrections to $g_A^0$ is given in \cite{Brodsky}.

\item Some authors, {\it e.g.}, \cite{TPCheng90,Geng}, in early 90's had tried to relate  $\la p|G\tilde G|p\ra$ with the gluon spin component $\Delta G$ of the proton; more precisely,
\be \label{eq:Gluonspin}
2m_N\Delta G\,\bar ui\gamma_5u=\la p|G\tilde G|p\ra \ .
\en
A measurement of $\Delta G$ will then allow us to extract $E_q$'s.  However, the above relation is repudiated on the ground that there is no twist-2 gauge-invariant local operator definition for $\Delta G$.
Another serious problem is that Eq. (\ref{eq:Gluonspin}) will imply a large isospin violation for the gluon spin component $\Delta G$; that is, $(\Delta G)_n$ is very different from $(\Delta G)_p$ since $\la N|G\tilde G|N\ra$ violates isospin symmetry \cite{Gross,Castro}.

\item It is known that in the presence of strong {\it CP} violation, the {\it CP}-odd operator $-(g^2/32\pi^2)\theta G\tilde G$ in QCD can be rotated away in terms of quark pseudoscalar densities, {\it i.e.}, the well-known Baluni operator \cite{Baluni:1978rf}
\be \label{eq:Baluni}
\delta{\cal L}_{CP}^{\rm Baluni}=\theta \bar m(\bar ui\gamma_5u+\bar di\gamma_5d+\bar si\gamma_5 s)
\en
with $\bar m=(1/m_u+1/m_d+1/m_s)^{-1}$. Naively, it is expected that the relation
\be \label{eq:Ei&GG}
\bar m(E_u+E_d+E_s)=
-\left< N \left| \frac{\alpha_s}{8\pi}G\tilde G
\right| N \right>
\en
permits one to recast all the unknown matrix elements in terms of the couplings $g_A^a$.  However, it is straightforward to check that Eqs.~(\ref{eq:gAEi}) and (\ref{eq:Ei&GG}) when combined together do not lead to any solution for $E_q$'s.  As pointed out in \cite{Cheng:1991}, a key to this paradox is the observation made in \cite{Aoki} that it is the disconnected insertion of $\delta{\cal L}_{CP}^{\rm Baluni}$ that is related to the insertion of $\theta G\tilde G$ in the hadronic process while the connected insertion of $\delta{\cal L}_{CP}^{\rm Baluni}$ must vanish for on-shell amplitudes.
Hence, if care is not taken, the use of Baluni's Lagrangian may lead to fake results, such as the electric dipole moments of the constituent quarks \cite{Abada}.

\item Another piece of information on $\la N|G\tilde G|N\ra$ comes from the U(1) Goldberger-Treiman (GT) relation valid in the chiral limit (for a review of the isosinglet GT relation, see \cite{Cheng1996}), from which one derives
\be
\left< N \left|{\alpha_s\over 8\pi}G\tilde G
\right| N \right>={f_\pi\over 2\sqrt{3}} g_{\eta_0 \! N\!N}+\cdots \ .
\en
In principle one can use this GT relation to fix the matrix element $\la N|G\tilde G|n\ra$ and to cross-check the result of Eq.~(\ref{eq:GGtilde}) given below.  In practice, this is not an easy task because of the mixing of $\eta_0$ with $\eta_8$ and $\pi^0$ and possibly the pseudoscalar glueball and the ill-determined couplings $g_{\eta' N\!N}$ and $g_{\eta N\!N}$.

\end{enumerate}

To summarize, we have listed several different methods of extracting the proton matrix elements of quark pseudoscalar densities and $G \tilde G$.  It appears that only the approach based on the large-$N_c$ chiral limit relation Eq.~(\ref{eq:sum}) still remains viable.  Let us first quote the results from \cite{Cheng1989}:
\be \label{eq:Eqs}
E_u &=& {1\over 2}{m_N\over m_dm_s(1+z+w)}\left[(m_d+2m_s)g_A^3+m_d\, g_A^8\right], \non \\
E_d &=& {1\over 2}{m_N\over m_dm_s(1+z+w)}\left[-(m_u+2m_s)g_A^3+m_u g_A^8\right],  \\
E_s &=& {1\over 2}{m_N\over m_dm_s(1+z+w)}\left[(m_u-m_d)g_A^3-(m_u+m_d) g_A^8 \right], \non
\en
and
\be \label{eq:GGtilde}
\left< N \left|{\alpha_s\over 8\pi}G\tilde G
\right| N \right> &=& {1\over 2}{m_N\over m_dm_s(1+z+w)}\Bigg\{ {2\over 3}m_dm_s(1+z+w)g_A^0+ m_s(m_d-m_u)g_A^3 \non \\
&& \quad +{1\over 3}[m_s(m_u+m_d)-2m_um_d]g_A^8 \Bigg\}
\en
with $z\equiv m_u/m_d$ and $w\equiv m_u/m_s$.  Numerically, $z=0.5$ and $w=0.025$ in this work.  The pseudoscalar-nucleon coupling now has the expression
\be \label{eq:sigmapp}
g_{\sigma\! N\!N} &=& {1\over 2}(\sqrt{2}G_F)^{1/2}m_N\Bigg\{ -{2\over 3}(\sum_h \xi_h)g_A^0+ \left[ \xi_u-\xi_d-(\sum_q \xi_q){1-z\over 1+z+w}\right]g_A^3  \non \\
&& +{1\over 3}\left[\xi_u+\xi_d-2\xi_s-(\sum_q \xi_q){1+z-2w\over 1+z+w}\right]g_A^8 \Bigg\}.
\en
In terms of $\Delta q$'s and $\bar m$ defined in Eq.~(\ref{eq:Baluni}), the matrix element $\la N |(\alpha_s/ 8\pi)G\tilde G| N \ra$ and the pseudoscalar-nucleon coupling can be recast to more symmetric forms
\be \label{eq:GGtilde2}
\left< N \left|{\alpha_s\over 8\pi}G\tilde G
\right| N \right> &=& {m_N\over m_dm_s(1+z+w)}(m_dm_s\Delta u+m_um_s\Delta d+m_um_d\Delta s) \non
\\
&=& {m_N\, \bar m} \left(\frac{\Delta u}{m_u} + \frac{\Delta d}{m_d} + \frac{\Delta s}{m_s} \right)
\en
and
\be \label{eq:g Deltaq}
g_{\sigma\! N\!N}&=&(\sqrt{2}G_F)^{1/2}m_N\Bigg\{ \left[\xi_u-(\sum_q\xi_q){1\over 1+z+w}\right]\Delta u+\left[\xi_d-(\sum_q\xi_q){z\over 1+z+w}\right]\Delta d \non \\
&& \qquad +\left[\xi_s-(\sum_q\xi_q){w\over 1+z+w}\right]\Delta s \Bigg\}
\\
&=& (\sqrt{2}G_F)^{1/2}m_N\Bigg\{
  \left[\xi_u - \frac{\bar m}{m_u}\sum_q\xi_q \right]\Delta u
+ \left[\xi_d - \frac{\bar m}{m_d}\sum_q\xi_q \right]\Delta d
+ \left[\xi_s - \frac{\bar m}{m_s}\sum_q\xi_q \right]\Delta s \Bigg\},
\non
\en
where use has been made of Eq.~(\ref{eq:gA Deltaq}).  Although the relation (\ref{eq:sum}) is subject to $1/N_c$ and chiral corrections, it is pertinent to regard Eqs.~(\ref{eq:Eqs})--(\ref{eq:g Deltaq}) as benchmark results.

It is interesting to notice that the general axion-nucleon coupling has precisely the same expression as Eq.~(\ref{eq:sigmapp}) or (\ref{eq:g Deltaq}) (see \cite{Srednicki} and Eq.~(9) on p.~497 of \cite{PDG}). However, the derivation of the axion coupling with the nucleons does not rely on the assumption of the large-$N_c$ chiral limit. This has to do with the fact that the physical axion does not have an anomalous $G\tilde G$ interactions \cite{Cheng1989}.

Before proceeding further, we would like to stress that contrary to the case for scalar-nucleon couplings, $g_{\sigma\! N\!N}$ does receive significant contributions from light quarks even in the chiral limit (see Eq. (\ref{eq:sigmapp}) or (\ref{eq:g Deltaq})). This is because the pseudoscalar boson has an admixture with $\pi,\eta_8$ and $\eta_0$ which in turn couple strongly with the nucleon. The smallness of the pseudoscalar-nucleon coupling is compensated by the smallness of the Goldstone bosons as the ratio of $m_q/m_\pi$, for example, remains finite in the chiral limit.

As for the couplings $g_A^a$, we note that since there is no anomalous dimension associated with the axial-vector currents $A_\mu^3$ and $A_\mu^8$, the non-singlet couplings $g_A^3$ and $g_A^8$ do not evolve with $Q^2$ and hence can be determined at $Q^2=0$ from low-energy neutron and hyperon beta decays.  Under flavor SU(3) symmetry, the non-singlet couplings are related to the SU(3) parameters $F$ and $D$ by
\be
g_A^3=\,F+D ~,\qquad g_A^8=\,3F-D ~.
\en
We use the updated coupling $g_A^3=1.2701\pm 0.0025$ \cite{PDG} and the value $F/D=0.575\pm0.016$ \cite{Close} to obtain
\be
F=0.464\pm 0.014\,,\qquad D=\,0.806\pm 0.008\,,\qquad g_A^8=0.585\pm0.039\,.
\en
Using this value of $g_A^8$, the recent measurements of $\Gamma_1^p$ by COMPASS \cite{Compass} and HERMES \cite{Hermes} imply that
\be
\Delta\Sigma_{_{\rm COMPASS}}(3\,{\rm GeV}^2) & = & 0.35\pm0.03\pm0.05\,, \non \\
\Delta\Sigma_{_{\rm HERMES}}(5\,{\rm GeV}^2) & = & 0.330\pm0.011({\rm theo})\pm0.025({\rm exp})\pm0.028({\rm evol})~.
\en
Therefore, $g_A^0$ is in the vicinity of 0.34\,. However, whether the flavor SU(3) symmetry can be applied to determine $g_A^8$ still remains controversial \cite{Jaffe,Leader,Karliner}. Indeed, there had been some suggestions that SU(3) symmetry might be badly broken and the errors could be as large as 25\% \cite{Jaffe,Savage}.  A recent calculation based on the cloudy bag model showed that $g_A^8=0.46\pm0.05$ which was reduced by as much as 20\% below the usual phenomenological value \cite{Bass}.   It is clear from Eq.~(\ref{eq:g1p}) that a decrease in $g_A^8$ will increase $g_A^0$ so that the magnitude of $\Delta s$ inferred from $-\Delta s={1\over 3}(g_A^8-g_A^0)$ is reduced from 0.08 to 0.03 (see Table~\ref{tab:pseudo}).  Interestingly, a very recent lattice calculation yields $\Delta s=-0.020\pm0.010\pm0.004$ at $\mu^2=7.4$ GeV$^2$ \cite{QCDSF:Deltas}.

There are still two remaining issues needed to be addressed.  First, Eqs.~(\ref{eq:Eqs})--(\ref{eq:g Deltaq}) are derived based on the relation (\ref{eq:sum1}) valid in the chiral and large-$N_c$ limits.  It is natural to ask how sizable the chiral and $1/N_c$ corrections to Eq.~(\ref{eq:sum1}) are.  In principle, this can be studied in the effective chiral Lagrangian approach that accommodates both corrections.  A detailed investigation will be reported elsewhere.  Second, since $\Delta q$'s and $g_A^0$ are thus far evaluated at $Q^2$ around 4 GeV
$^2$, how does one evolve them to $Q^2=0$?  Unfortunately, the evolution into $Q^2=0$ is beyond the perturbative QCD regime and hence not calculable in a perturbative way.

Finally we come to the estimation of the coupling strength of $g_{\sigma\! N\!N}$.  We shall assume that $\xi_q=1$ for definiteness and simplicity and evaluate various parameters for two cases: (i) $g_A^8=0.585$ and $g_A^0=0.34$ and (ii) $g_A^8=0.46$ and $g_A^0=0.37$\,.~\footnote{The value $g_A^0=0.37$ is obtained from Eq.~(\ref{eq:g1p}) with the input of $g_A^8=0.46$\,.}
Note that for the neutron, $(g_A^3)_n$ is opposite to $(g_A^3)_p$; that is, $(g_A^3)_n=-(F+D)$.  The numerical results are summarized in Table~\ref{tab:pseudo}.  Obviously, the nucleon matrix elements $\la N|\bar si\gamma_5s|N\ra$ and $\la N|G\tilde G|N\ra$ strongly violate isospin symmetry, as pointed out in \cite{Gross,Castro}.  For example, it is shown in \cite{Castro} that $A_\mu^8$ can be an isoscalar operator only if the pseudoscalar operator $\bar si\gamma_5s$ is not an isoscalar one.  It is worth mentioning that the result for $\la N|G\tilde G|N\ra$                                                                         in Table~\ref{tab:pseudo} will be useful if one considers the gluon field coupling to the fermionic dark matter through the effective operator $(\bar\chi\gamma^5\chi)G\tilde G$ or to the complex scalar dark matter described by the interaction $(\chi^\dagger\chi)G\tilde G$.

\begin{table}[t]
\caption{Summary of parameters.  All are in units of MeV except for the dimensionless $\Delta q$'s and $g_{\sigma\! N\!N}$. The values in the upper (lower) sector are obtained using $g_A^8=0.586$ and $g_A^0=0.34$ ($g_A^8=0.46$ and $g_A^0=0.37$). Note that $g_A^3$ is equal to $1.2701$ for the proton and $-1.2701$ for the neutron. The assumptions of three heavy quarks and $\xi_q=1$ have been made. The quark spin components $\Delta q$'s and the coupling $g_A^0$ are evaluated at $Q^2$ of order 4 GeV${^2}$.
}
\label{tab:pseudo}
\begin{center}
\begin{tabular}{c c c c c c c c c}
\hline\hline
$N$~~~& $m_u E_u$~~~ & $m_d E_d$~~~ & $m_s E_s$~~~ & $\la N|{\alpha_s\over 8\pi}G\tilde G|N\ra$~~~ & $\Delta u$~~~ & $\Delta d$~~~ & $\Delta s$~~~ & $g_{\sigma\! N\!N}$  \\ \hline
$p$~~~ & 405~~~ & $-787$~~~ & $-466$~~~ & 389 & 0.85~~~ & $-0.42$~~~ & $-0.08$~~~ & $-8.2\times 10^{-3}$ \\
$n$~~~ & $-396$~~~ & $796$~~~ & $-75$~~~ & $-2$ & $-0.42$~~~ & 0.85~~~ & $-0.08$~~~ & $1.3\times 10^{-3}$ \\
\hline
$p$~~~ & 404~~~ & $-788$~~~ & $-408$~~~ & 380 & 0.84~~~ & $-0.44$~~~ & $-0.03$~~~ & $-7.8\times 10^{-3}$ \\
$n$~~~ & $-397$~~~ & $795$~~~ & $-17$~~~ & $-11$ & $-0.44$~~~ & 0.84~~~ & $-0.03$~~~ & $1.7\times 10^{-3}$ \\
\hline\hline
\end{tabular}
\end{center}
\end{table}

%
\section{Summary \label{sec:summary}}
%

The couplings between nucleons and scalar or pseudoscalar Higgs bosons are of high interest,
particularly for the estimation of cross sections between dark matter and nuclei.  Such calculations rely on sufficient knowledge of various nucleon matrix elements of quark scalar and pseudoscalar densities.  In view of better understanding of baryon masses, quark spin components and lattice studies of the pion-nucleon and quark sigma terms in recent years, we update the scalar/pseudoscalar couplings of the nucleons in this work.

Early calculations usually find a larger strange quark content, characterized by the parameter $y$, and a larger value of the strange-quark sigma term $\sigma_s$.  They are partly due to a larger $\sigma_{\pi\! N} \simeq 55- 60$ MeV, assumed to be the on-shell pion-nucleon scattering amplitude at the Cheng-Dashen point, as well as a smaller $\sigma_0 \simeq 24$ MeV, extracted from the baryon masses using the leading-order formulae in quark masses.  In this work, we take the cue from recent lattice studies that $\sigma_s$ is much smaller than previously thought. For definiteness, we use $\sigma_s=50$ MeV as a benchmark value, which is consistent with the global fit of the lattice calculations.  This together with $\sigma_0 \simeq 36$ MeV obtained using the baryon mass formulae computed to the second order in light quark masses leads to $\sigma_{\pi\! N} \simeq 40$ MeV in good agreement with recent lattice determinations \cite{Durr,Bali:ms}.  Therefore, the strange quark content in the nucleon is $y \simeq 0.09$\,.  This is roughly a factor of two smaller than the previous estimate of $y\simeq 0.2$\,.  The relative importance of the light quarks in the effective scalar Higgs-nucleon coupling is shown by the ratios $\sigma_u : \sigma_d : \sigma_s \simeq 1 : 1.3 : 3$, to be contrasted with $1 : 1.5 : 6$ in \cite{Gondolo} and $1 : 1.3 : 6$ in \cite{Ellis:2000ds}.  (See Appendix~\ref{sec:comparison} for a detailed comparison.)  We find that the coupling $g_{\phi pp} \simeq 1.1 \times 10^{-3}$, of which the strange quark contribution is $18\%$, a factor of two smaller than what is usually used in the calculations of direct detection of dark matter \cite{Gondolo,Ellis:2000ds}.  The dominant contribution ($\sim 70\%$) comes from the heavy quarks.  Moreover, the ratio $g_{\phi pp}/g_{\phi nn}$ is very close to unity, showing essentially no isospin violation.

For the pseudoscalar Higgs couplings with the nucleons, several different methods for extracting the nucleon matrix elements of pseudoscalar quark densities and $G\tilde G$ are reviewed.  We conclude that the method based on the relation $\la N|\bar ui\gamma_5u+\bar di\gamma_5d+\bar si\gamma_5s|N\ra=0$ derived in the chiral and large-$N_c$ limits still remains viable.  We then take two sets of axial coupling constants $(g_A^3,g_A^8,g_A^0) = (1.2701,0.585,0.34)$ and $(1.2701,0.46,0.37)$ to compute the corresponding quark spin
components $\Delta q$'s.  The second set of parameters is motivated by the recent observation that there is a possible SU(3)-breaking effect on the determination of $g_A^8$ implied by both model and lattice analyses.  For example, a recent calculation based on the cloudy bag model finds $g_A^8$ to be reduced by $\sim 20\%$ from the usual phenomenological value in the first set.  We find that $-\Delta s$ changes from $0.08$ to $0.03$ due to such a reduction in $g_A^8$.  The isospin violation in the nucleon matrix elements $\la N|\bar si\gamma_5s|N\ra$ and $\la N|G\tilde G|N\ra$ results in significantly different $g_{\sigma\! N\!N}$ for $N = p$ and $n$.  We obtain $g_{\sigma nn}/g_{\sigma pp} \simeq -0.16$ and $-0.22$ for the two parameter sets, respectively.

%
\section*{Acknowledgments}
%

The authors would like to thank T.~C.~Yuan for bringing our attention to the DarkSUSY package and Jose Manuel Alarcon for useful discussions.
This research was supported in part by the National Science Council of Taiwan, R.~O.~C. under Grant Nos.~NSC-100-2112-M-001-009-MY3 and NSC-100-2628-M-008-003-MY4.

%
\appendix
%

%
\section{Comparison with other works \label{sec:comparison}}
%

In the lattice and dark matter communities, it is conventional to consider the quantities defined by
\be
f_{T_q}\equiv {\la N|m_q\bar qq|N\ra\over m_N}={\sigma_q\over m_N}, \qquad f_{T_G}=1-\sum_q^{u,d,s}f_{T_q} \ ,
\en
parameterizing the fractions of the nucleon mass $m_N$ carried by the corresponding quarks and gluons.
In terms of these parameters, the coupling $g_{\phi pp}$ (see Eq. (\ref{eq:phipp})) has the expression
\be
g_{\phi pp} = (\sqrt{2}G_F)^{1/2}m_N\left(\sum_\ell \zeta_\ell \, f_{T_\ell}+{2\over 27}f_{T_G}\sum_h\zeta_h\right).
\en
A popular set of values is that used in the DarkSUSY package \cite{Gondolo}
\be
&& f_{T_u}^{(p)}=0.023\,, \quad f_{T_d}^{(p)}=0.034\,, \quad f_{T_s}^{(p)}=0.14\,, \quad f_{T_{c,b,t}}^{(p)}=0.0595\,, \non \\
&& f_{T_u}^{(n)}=0.019\,, \quad f_{T_d}^{(n)}=0.041\,, \quad f_{T_s}^{(n)}=0.14\,, \quad f_{T_{c,b,t}}^{(n)}=0.0592\,,
\en
and
\be
\Delta u^{(p)}=\Delta d^{(n)}=0.77\,, \qquad \Delta d^{(p)}=\Delta u^{(n)}=-0.40\,, \qquad \Delta s^{(p)}=\Delta s^{(n)}=-0.12\,.
\en
Another set of parameters commonly quoted in the literature is \cite{Ellis:2000ds}
\be
&& f_{T_u}^{(p)}=0.020\,, \quad f_{T_d}^{(p)}=0.026\,, \quad f_{T_s}^{(p)}=0.118\,, \quad f_{T_{c,b,t}}^{(p)}=0.0619\,, \non \\
&& f_{T_u}^{(n)}=0.014\,, \quad f_{T_d}^{(n)}=0.036\,, \quad f_{T_s}^{(n)}=0.118\,, \quad f_{T_{c,b,t}}^{(n)}=0.0616\,,
\en
and
\be
\Delta u^{(p)}=\Delta d^{(n)}=0.78\,, \qquad \Delta d^{(p)}=\Delta u^{(n)}=-0.48\,, \qquad \Delta s^{(p)}=\Delta s^{(n)}=-0.15\,,
\en
where only the central values are listed.

For a comparison, in this work we have
\be
&& f_{T_u}^{(p)}=0.017\,, \quad f_{T_d}^{(p)}=0.023\,, \quad f_{T_s}^{(p)}=0.053\,, \quad f_{T_{c,b,t}}^{(p)}=0.0672\,, \non \\
&& f_{T_u}^{(n)}=0.012\,, \quad f_{T_d}^{(n)}=0.033\,, \quad f_{T_s}^{(n)}=0.053\,, \quad f_{T_{c,b,t}}^{(n)}=0.0668\,,
\en
for the choice of $\sigma_s=50$ MeV, while
\be
\Delta u^{(p)}=\Delta d^{(n)}=0.85\,, \qquad \Delta d^{(p)}=\Delta u^{(n)}=-0.42\,, \qquad \Delta s^{(p)}=\Delta s^{(n)}=-0.08
\en
for $g_A^8=0.585$ and
\be
\Delta u^{(p)}=\Delta d^{(n)}=0.84\,, \qquad \Delta d^{(p)}=\Delta u^{(n)}=-0.44\,, \qquad \Delta s^{(p)}=\Delta s^{(n)}=-0.03
\en
for $g_A^8=0.46$\,.

It is noticed that in the scalar Higgs-nucleon couplings, the contributions from light quarks in our work are generally smaller than those used in \cite{Gondolo,Ellis:2000ds}.  In particular, the strange quark contribution is nearly two to three times smaller.  The heavy quarks have a more dominant effect.  The strange-quark spin component is also found to be 2 to
5 times smaller in magnitude than those in \cite{Gondolo,Ellis:2000ds}.


\end{document}